# Inadequate experimental methods and erroneous epilepsy diagnostic criteria result in confounding acquired focal epilepsy with genetic absence epilepsy

**Authors:** Raimondo D'Ambrosio, Ph.D., Clifford L. Eastman, Ph.D., and John W. Miller, M.D., Ph.D.

**Affiliation:** Department of Neurological Surgery, Regional Epilepsy Center, University of Washington, Seattle, WA 98104, USA.

**Summary**

Here we provide a thorough discussion of the study conducted by Rodgers et al. (2015) to investigate focal seizures and acquired epileptogenesis induced by head injury in the rat. This manuscript serves as supplementary document for our letter to the Editor to appear in the Journal of Neuroscience. We find that the subject article suffers from poor experimental design, very selective consideration of antecedent literature, and application of inappropriate epilepsy diagnostic criteria that, together, lead to unwarranted conclusions.

**Introduction**

Many laboratories have been developing new etiologically realistic models to match different epilepsy syndromes in humans, with the hope that such models will reveal mechanisms that are actually relevant to the corresponding human epilepsies and identify treatments that will successfully translate to the corresponding patient populations. Development of these models, however, requires mastery of new techniques to produce epilepsy, and a good understanding of the corresponding human syndromes and of the established clinical practices used to evaluate such conditions. In this case, the clinical disorder is posttraumatic epilepsy (PTE), and the techniques used are fluid percussion injury (FPI), a realistic model of head injury in the rat, and electrocorticography (ECoG). FPI is a robust, yet complex technique that has been used and refined by the traumatic brain injury community since the 70s. It is recognized as a useful realistic model of human contusive closed head injury (Thompson et al., 2005) and it was used to demonstrate that rats, like humans, could develop posttraumatic epilepsy after a single contusive event (D'Ambrosio et al., 2004). Human PTE is an epilepsy disorder that is induced by head injury, with contusion being a potent risk factor. In PTE patients, convulsive or nonconvulsive seizures most often arise from the frontal neocortex and the temporal lobe (Gupta et al., 2014; Diaz-Arrastia et al., 2009). Currently, no clinical treatment exists to prevent the development of PTE in head injury patients, and, in 30-40% of patients with PTE the seizures cannot be controlled with any of the available antiepileptic medications. The bulk of these drug-resistant seizures are focal and non-convulsive, while the convulsive type is generally easier to control with current medications (Semah et al., 1998). In addition, mesial temporal epilepsy is often very successfully treated by surgical resection, while the neocortical variant has a lower surgical success rate (Jeha et al., 2007; Mosewich et al., 2000). Thus, one of the current translational imperatives is to develop good animal models of *focal non-convulsive seizures* to help develop new treatments for epilepsy patients. Several independent laboratories have successfully deployed FPI, and induced various forms of convulsive and non-convulsive epileptic seizures. Our laboratory has focused on developing, characterizing and optimizing a rat model of focal neocortical non-convulsive PTE induced by rostral parasagittal FPI (rpFPI; Curia et al., 2011; D'Ambrosio et al., 2004; 2005; 2009; Eastman et al., 2015), and deploying it to test treatments (Eastman et al., 2010; 2011; D'Ambrosio et al., 2013; Curia et al., 2015).

**Discussion**

In this study, Rodgers et al. attempt to follow up on this work. They exposed rats to an unconventional FPI protocol, and found no acquired focal seizures, convulsive or non-convulsive. They instead recorded frequent bilateral spontaneous spike and wave discharges (SWDs) that were equally distributed among injured and uninjured rats. The authors, erroneously claim that their bilateral SWDs are identical to the acquired focal neocortical epileptic seizures described by our laboratory after rpFPI, and argue that these results raise concern about the translational relevance of the rat FPI-PTE model. To the contrary, these investigators: 1) used an unconventional FPI method that simply did not induce an epileptogenic injury, 2) used rats that appear to suffer from an unprecedented high incidence of genetic absence epilepsy (bilateral SWDs) never before reported in young male Sprague Dawley rats, 3) failed to use the correct diagnostic criteria used clinically to distinguish focal epilepsy from genetic absence epilepsy, and 4) systematically failed to acknowledge or discuss a large body of published data that would invalidate their conclusion. As a result the authors compared absence epilepsy in their control rats with the same absence epilepsy in their FPI rats, and concluded that they look similar. Thus, contrary to the authors' assertion, the study has no implication for the translational relevance of the FPI model of PTE, or for any of the other etiologically realistic acquired epilepsy models that are being used in a growing number of laboratories and that induce non-convulsive seizures.

Some of our most important concerns about this paper are detailed below:

First, the FPI methods used in this study are inadequate. While this work seems to be presented as a meticulous and faithful replication of our rpFPI-PTE model, it is actually impossible to determine how FPI was administered. The authors do not disclose the device used to administer FPI, and describe their methods *only* by citation of prior studies that used two very different FPI devices and methodologies (D'Ambrosio et al., 2004; Frey et al., 2009; Rodgers et al., 2012), as if these disparate methods were interchangeable. They are not. Two issues are key: the FPI device, and the fluid path connecting the device to the animal. Based on previous correspondence with the senior author, and considering his past publications, we understand that they inflicted FPI with a device where the mechanical energy is generated by an electronically-controlled picospritzer (Frey et al., 2009; Rodgers et al., 2012). Conversely, *all* laboratories that have successfully induced PTE with FPI used the classic FPI device which consists of a saline-filled cylinder from which saline is ejected into the skull cavity onto the dura to create a contusion when a weighted pendulum strikes a piston (Thompson et al., 2005; D'Ambrosio et al., 2004; Kharatishvili et al., 2006; Goodrich et al., 2013; Shultz et al., 2013). This is critical because *no protocol to induce epileptogenesis with a picospritzer has* *ever been described* and there is, to date, no evidence that picospritzer-based FPI can deliver the kinetic energy required to properly injure the brain to induce any of the PTE syndromes (convulsive or non-convulsive; neocortical or mesial temporal) that have been reported after conventional FPI by several laboratories.

The authors' attempt to match pressure pulse amplitude and duration is inadequate to establish equivalency of injury performed by other laboratories, and no data on the actual kinetic energy delivered, or on the volume of injected saline, or on the contusion induced by their picospritzer-based method is presented in the paper. The proxy measure of the injuring kinetic energy -the pressure pulse waveform- is measured in all laboratories by a transducer located inside the FPI apparatus several inches from the dura. Thus, the kinetic energy actually delivered to the dura, the volume of saline injected onto the dura in the skull, and the induced brain compression and contusion, are all critically influenced by the gauge, length and flexibility of the fluid path from the transducer to the dura. Experienced laboratories manage all of these paramenters (and many more) to ensure the appropriate delivery of the required kinetic energy to the dura (Dixon et al., 1988; McIntosh et al., 1989). The parameters of the pressure pulse provided in the FPI literature are used by experienced laboratories only to compare injuries administered using strictly identical FPI devices and methods. The fluid path typically differs in different instruments (even of the same type), and can even be affected by the well-known problem of compressible air bubbles that must be purged from the system for reproducibility and reliability. Even small differences in the fluid path between the pressure transducer and the dura can cause significant changes in pressure that would not be detected by the pressure transducer. The authors' use of acute mortality rate and righting time also does not establish equivalency of injury. It has been demonstrated that these acute measures are inadequate indicators of the epileptogenic potential of classic FPI, even when all other variables are held constant within the same laboratory (Curia et al., 2011). We have shown that, all other things being equal, reducing the amplitude of the FPI pressure pulse from 3.4 atm to 2 atm results in a dramatically lower incidence and frequency of seizures, in fact, with no seizures in the early months post-injury, without appreciable change in mortality rate, posttraumatic apnea or righting times (Curia et al., 2011). In addition, the pathological evidence presented in this paper is scant (one cresyl-violet stained section per group), and is inadequate to indicate equivalency of injury induced by other laboratories, especially in view of the fact that the PTE induced by conventional FPI appears to be independent of major structural, functional, and behavioral changes induced by TBI (Shultz et al., 2013; Curia et al., 2011).

Thus, there is no basis for confidence that the FPI methods in this report were similar to the conventionally delivered FPI that induce epileptogenesis in other laboratories. Indeed, the failure to induce PTE in this study demonstrates the injuries *were different*. There are many parameters in experimental brain injury models like the FPI that must be properly controlled to obtain reproducible results. Before being used for epilepsy research, novel implementations of FPI must first be proven to produce the same results as the classic method does (Hameed et al., 2014). Explicit comparison to conventional methods that have been proven to work is especially relevant in a report, such as this one, which highlights negative findings -  particularly given that no protocol to induce PTE with a picospritzer has *ever* been published. Indeed, the authors have not yet demonstrated that, using their picospritzer-based FPI methods, they can replicate tonic-clonic convulsions reported after lateral FPI (Kharatishvili et al., 2006; Shultz et al., 2013), which they regard as a bona fide PTE model.

Second, Rodgers et al report of a high incidence (70% - 100%) of very frequent (~100/day) bilateral SWDs in uninjured young male Sprague Dawley rats, is unprecedented and puzzling. While the occurrence of age-dependent bilateral SWDs in most laboratory rat strains is well documented, *all* prior studies we are aware of have found them to be absent or very rare in young (<3-5 months) male Sprague Dawley rats (Willmore et al., 1978a; 1978b; Aldinio et al., 1985; Aporti et al., 1986; Buzsaki et al., 1990; Willoughby and MacKenzie, 1992; D'Ambrosio et al., 2005; Pearce et al., 2014). Despite citing several of these studies, Rodgers et al fail to note or discuss this critical discrepancy between their data and the literature, and leave their aberrant findings completely unexplained. These findings could reflect a high rate of false-positive detections due to poor accuracy of the unpublished and unvalidated methods they used to automate the quantification of SWDs. In fact, the authors do not provide enough details of the algorithm, the training dataset, and of the process of identification of false positives to judge the quality of their approach. Another plausible explanation for the unprecedented SWD-activity reported in this study is that the Sprague Dawley rats they obtained from Harlan Laboratories differed genetically from those used in the laboratories that found few or no SWDs in the young male Sprague Dawley rats, and were more prone to SWDs. Recent work has shown that genetic drift can affect the behavioral performance of Sprague Dawley rats obtained from different vendors, and even from different colonies of the same vendor (Fitzpatrick et al., 2013). Genetic drift has been shown to greatly affect frequency and incidence of bilateral SWDs in GAERS (Powell et al., 2014). Whatever the explanation, it seems unlikely that the frequent occurrence of SWDs in virtually all young male Sprague Dawley rats could have eluded detection in *six independent laboratories* over the span of almost 40 years.

Third, the authors do not use the correct diagnostic criteria used in humans to distinguish between the seizures of focal neocortical epilepsy and those of generalized absence epilepsy, but dwell instead on the morphology and power spectrum of the ECoG discharges, which *are not* clinically recognized as adequate diagnostic criteria to differentiate these seizure types. Indeed, clinical neurophysiology has long recognized that certain waveform features, like theta frequency activity, with or without fast spikes or spike-wave complexes, can also be found in focal neocortical epilepsy, especially frontal and temporal, and of any etiology, including trauma (Niedermeyer et al., 1970; Hughes, 1980). Because of this, the clinically recognized criteria for determination of focal seizures are the consistent origin from a specific pathological anatomical area (the epileptic focus), the lateralization of the electrical seizure, and the delayed spread, if it occurs (Stefan et al., 1997; Muro and Connolly., 2014). A matching brain lesion, if present, will support the diagnosis. These features are diagnostic, because human focal seizures initiate consistently from the same pathological anatomical area (the epileptic focus), and spread with a variable time lag from about 1 second to tens of seconds (Gotz-Trabert et al., 2008). Conversely, initiation of a seizure from both hemispheres (i.e. electrical bilaterality within a very short time frame), is a key clinical diagnostic criterion used to distinguish genetic absence epilepsy from focal seizures in humans (Drury and Henry, 1993). The authors, report 80%-90% of their SWDs to be bilateral and begin synchronously within 100ms in all channels, and precipitate from both frontal lobes. They also report that the remaining SWDs exhibit no consistent lateralization, i.e. they initiate from either frontal lobe. In fact, these observations *demonstrate* that they are not dealing with an anatomical epileptic focus but rather with absence epilepsy. The authors' claim that absence epilepsy can be "focal" is misleading in the context of clinical differential diagnosis from acquired epilepsy, because absence epilepsy does not have a pathological epileptic focus like focal seizures do. The electrical discharges in absence epilepsy do not localize to a pathological area, exhibit no consistent lateralization, and initiate from either frontal lobe even when a sophisticated phase analysis is conducted to detect the earliest (within milliseconds) network precipitating it in both humans (Holmes, 2008) and rat (Meeren et al., 2002).

In stark contrast with the seizures reported by Rodgers et al., our ECoG recordings demonstrate that most seizures induced by rpFPI have a consistent unilateral focal perilesional onset, *with or without* subsequent spread to distal cortical areas. The vast majority of the seizures that we observe within few months after FPI are "isolated" (grade 1 in our terminology). These are detected only at the perilesional electrode next to the epileptic focus, and range in duration from 1s to over a minute. When the focal seizure spread (grade 2 in our terminology) the spread occurs with a time-lag much longer than the authors' 100ms criterion, ranging from about 1 to tens of seconds, and consistent with the time lag of the spread of human focal neocortical seizures (Gotz-Trabert et al., 2008). The perilesional focality and spread of the rpFPI-induced seizures has been demonstrated in several of our papers (For example, Figs. 1 and 4B in D'Ambrosio et al., 2004; Figs. 2A, 2B In D'Ambrosio et al., 2005; Figs. 2I, 2F, 2J, 7D in D'Ambrosio et al., 2009; Figs. 10E, 10G in Eastman et al., 2010; Fig. 6 in Curia et al., 2011; Fig. 7 in Eastman et al., 2011; Fig. 1 in Eastman et al., 2015), including two with grid recordings (Fig. 4 in D'Ambrosio et al., 2004; Figs. 2A-F in D'Ambrosio et al., 2009). In addition, consistent with their perilesional neocortical origin, we have shown in blind and randomized studies that rpFPI-induced focal seizures are potently prevented by a very focal treatment of the perilesional frontal neocortex (cooling by $2^oC$; D'Ambrosio et al., 2013). Thus, the focality and lateralization of the seizure induced by rpFPI, and the existence of the pathological epileptic focus originating them, is beyond question, and they distinguish these seizures from absence epilepsy and bilateral SWDs, *by clinically recognized criteria*.

Moreover, the authors err in asserting that it is established that short duration and spectral power at 7-9Hz is uniquely characteristic of rodent genetic bilateral SWDs. Many acquired epilepsy models, including cortical iron injection (Willmore et al., 1978a; 1978b), head injury (D'Ambrosio et al., 2004), photothrombic stroke (Paz et al., 2013), prolonged hippocampal electrical (Norwood et al., 2008) or kainate-mediated (Maroso et al., 2011; Roucard et al., 2014) excitation, and perinatal hypoxia (Rakhade et al., 2011) induce seizures that start with dominant spectral power in the theta band (4-9Hz). In fact, like the neocortical seizures described in the rpFPI model, the acquired focal seizures (both complex and simple partial) described in the neocortex of drug-resistant epilepsy patients can be short and present with a dominant spectral power in the theta band (Gotz-Trabert et al., 2008; Ikeda et al., 2009; D'Ambrosio et al., 2009; Butler et al., 2011). These frontal neocortical focal seizures described in humans often have the same electrophysiological properties, unilateral focality (with or without spread), spectral power, and even ictal behavioral arrest as those we observed in our rats with frontal neocortical focal seizures. The temporal neocortex has also been found to discharge with the same focal theta activity (Lüders et al., 1993). The readers can easily compare the electroclinical and spectral description of the human seizures shown in Figs. 5 and 7 in Ikeda et al., 2009, Fig. 1 in Butler et al., 2011, and also Fig. 1C in Gotz-Trabert et al., 2008, and Fig. 7 in Lüders et al., 1993, with the frontal neocortical seizures induced by rpFPI (e.g., Fig. 1 in D'Ambrosio et al., 2004; Fig. 2 in D'Ambrosio et al., 2009). Even the duration of the rpFPI-induced focal neocortical seizures is consistent with that found in human frontal neocortical epileptic foci with invasive monitoring. The readers can easily compare Fig. 1B in Eastman et al., 2015 and Fig. 2 in Curia et al., 2011, with Fig. 2A in Stead et al., 2010, and Fig. 7 in Ikeda et al., 2009. Thus, it is the epileptic discharge of the neocortex that often times, and independently from etiology, presents with brief events and a strong theta component of the spectrum in *both rodents and humans.*

Lastly, the authors failed to note that the posttraumatic epileptic seizures that develop after rpFPI also have pharmacology distinct from that of genetic bilateral SWDs in the rat. rpFPI-induced focal seizures were not controlled by carisbamate (Eastman et al., 2011) and, in 60% of the animals, by valproate (Eastman et al., 2010), which both potently controlled genetic bilateral SWDs (Francois et al., 2008; Marescaux et al., 1984; Dedeurwaerdere et al., 2011). In addition, as noted above, the ability to potently prevent rpFPI-induced focal seizures by mild focal cooling of just the perilesional neocortex demonstrates them to be acquired and originate from a genuine pathological epileptic focus (D'Ambrosio et al., 2013). All these results are impossible to reconcile with a claim that rpFPI-induced focal seizures are similar to genetic bilateral SWDs.

**Conclusions**

There is *no evidence* that the FPI administered in this study was adequate to induce epileptogenesis, nor that the authors' methods could induce any of the different forms of PTE, convulsive or non-convulsive, reported after FPI by many different laboratories. Except for the *unprecedented* and *unexplained* abundance of SWDs in young male Sprague Dawley rats used by the authors, this absence epilepsy is similar to that described in aging rats of this and other strains, and is clearly distinct from human focal neocortical epilepsy and from the rpFPI-induced focal neocortical seizures in the rat. Because the authors could not induce focal seizures by FPI, they *ended up comparing absence epilepsy in their controls with absence epilepsy in FPI rats*, and concluded that they look similar. They also used inappropriate epilepsy diagnostic criteria that cannot distinguish between focal non-convulsive seizures and genetic absence epilepsy. Moreover, the authors failed to consider all literature conflicting with their conclusion, and *surmised* similarities between the absence epilepsy in their rats with the focal seizures we induce by rpFPI. The absence-like SWDs described in this study have no obvious implication for what constitutes a model of acquired non-convulsive epilepsy, other than requiring investigators to be aware of their existence and to design their studies to recognize them and manage them.

To this end, the use of young animals with no or low background of SWD, like those used in many other laboratories, would simplify the work (Curia et al., 2015). In addition, the adoption in preclinical studies of clinically recognized epilepsy diagnostic criteria would facilitate the identification of new treatments that translate to the corresponding patient populations, especially when evaluating etiologically realistic models of acquired epilepsy that can present with human-like seizures (D'Ambrosio and Miller, 2010).

Conflict of interest: Drs. D'Ambrosio and Miller are co-founders of Therma Neurosciences Inc., a startup company tasked with developing a focal cooling device for the treatment of human drug-resistant focal epilepsy.

# REFERENCES

1. Aldinio C, Aporti F, Calderini G, Mazzari S, Zanotti A, Toffano G (1985) Experimental models of aging and quinolinic acid. Methods Find Exp Clin Pharmacol. 1985 Nov;7(11):563-8.
2. Aporti F, Borsato R, Calderini G, Rubini R, Toffano G, Zanotti A, Valzelli L, Goldstein L (1986) Age-dependent spontaneous EEG bursts in rats: effects of brain phosphatidyl-serine. Neurobiol Aging 7(2):115-20.
3. Butler T, Ichise M, Teich AF, Gerard E, Osborne J, French J, Devinsky O, Kuzniecky R, Gilliam F, Pervez F, Provenzano F, Goldsmith S, Vallabhajosula S, Stern E, Silbersweig D (2013) Imaging inflammation in a patient with epilepsy due to focal cortical dysplasia. J Neuroimaging 23(1):129-31.
4. Buzsáki G, Laszlovszky I, Lajtha A, Vadász C. (1990) Spike-and-wave neocortical patterns in rats: genetic and aminergic control. Neuroscience. 38(2):323-33.
5. Curia G, Levitt M, Fender JS, Miller JW, Ojemann J, D'Ambrosio R (2011) Impact of injury location and severity on posttraumatic epilepsy in the rat: role of frontal neocortex. Cereb Cortex 21(7):1574-92.
6. Curia G, Eastman CL, Miller JW, D'Ambrosio R (2015) Modeling posttraumatic epilepsy for therapy development. In: Gerald A. Grant and Daniel Laskowitz, Eds. Translational Research in Traumatic Brain Injury, Frontiers in Neuroscience, Taylor & Francis Book / CRC Press, Boca Raton, Fl. *in press*
7. D'Ambrosio R, Eastman CL, Darvas F, Fender JS, Verley DR, Farin FM, Wilkerson HW, Temkin NR, Miller JW, Ojemann J, Rothman SM, Smyth MD (2013) Mild passive focal cooling prevents epileptic seizures after head injury in rats. Ann Neurol 73(2):199-209.
8. D'Ambrosio R, Fairbanks JP, Fender JS, Born DE, Doyle DL, Miller JW (2004) Post-traumatic epilepsy following fluid percussion injury in the rat. Brain 127(Pt 2):304-14.
9. D'Ambrosio R, Fender JS, Fairbanks JP, Simon EA, Born DE, Doyle DL, Miller JW (2005) Progression from frontal-parietal to mesial-temporal epilepsy after fluid percussion injury in the rat. Brain 128(Pt 1):174-88.
10. D'Ambrosio R, Hakimian S, Stewart T, Verley DR, Fender JS, Eastman CL, Sheerin AH, Gupta P, Diaz-Arrastia R, Ojemann J, Miller JW (2009) Functional definition of seizure provides new insight into post-traumatic epileptogenesis. Brain 132(Pt 10):2805-21
11. D'Ambrosio R, Miller JW (2010) What is an Epileptic Seizure? Unifying Definitions in Clinical Practice and Animal Research to Develop Novel Treatments. Epilepsy Currents, 2010, 10(3):61-66. PMID: 20502593. Answer in Epilepsy Currents, 2010, 10(4):90. PMID: 20697503. PMCID 2873641
12. Dedeurwaerdere S, van Raay L, Morris MJ, Reed RC, Hogan RE, O'Brien TJ (2011) Fluctuating and constant valproate administration gives equivalent seizure control in rats with genetic and acquired epilepsy. Seizure 20(1):72-9. doi:10.1016/j.seizure.2010.10.011.
13. Diaz-Arrastia R, Agostini MA, Madden CJ, Van Ness PC (2009) Posttraumatic epilepsy: the endophenotypes of a human model of epileptogenesis. Epilepsia, 50 Suppl 2:14-20. doi: 10.1111/j.1528-1167.2008.02006.x. Review.
14. Dixon CE, Lighthall JW, Anderson TE Physiologic, histopathologic, and cineradiographic characterization of a new fluid-percussion model of experimental brain injury in the rat. J Neurotrauma (1988) 5(2):91-104.
15. Drury I, Henry TR (1993) Ictal patterns in generalized epilepsy. J Clin Neurophysiol. Jul;10(3):268-80.

16. Eastman CL, Fender JS, Temkin NR, D'Ambrosio R (2015;) Optimized methods for epilepsy therapy development using an etiologically realistic model of focal epilepsy in the rat. Exp Neurol 264:150-62.
17. Eastman CL, Verley DR, Fender JS, Stewart TH, Nov E, Curia G, D'Ambrosio R (2011) Antiepileptic and antiepileptogenic performance of carisbamate after head injury in the rat: blind and randomized studies. J Pharmacol Exp Ther 336(3):779-90.
18. Eastman CL, Verley DR, Fender JS, Temkin NR, D'Ambrosio R (2010) ECoG studies of valproate, carbamazepine and halothane in frontal-lobe epilepsy induced by head injury in the rat. Exp Neurol 224(2):369-88.
19. Fitzpatrick CJ, Gopalakrishnan S, Cogan ES, Yager LM, Meyer PJ, Lovic V, Saunders BT, Parker CC, Gonzales NM, Aryee E, Flagel SB, Palmer AA, Robinson TE, Morrow JD (2013) Variation in the form of Pavlovian conditioned approach behavior among outbred male Sprague-Dawley rats from different vendors and colonies: sign-tracking vs. goal-tracking. PLoS One 8(10):e75042. doi: 10.1371/journal.pone.0075042.
20. François J, Boehrer A, Nehlig A (2008) Effects of carisbamate (RWJ-333369) in two models of genetically determined generalized epilepsy, the GAERS and the audiogenic Wistar AS. Epilepsia 49:393-399.
21. Frey LC, Hellier J, Unkart C, Lepkin A, Howard A, Hasebroock K, Serkova N, Liang L, Patel M, Soltesz I, Staley K (2009) A novel apparatus for lateral fluid percussion injury in the rat. J Neurosci Methods 177(2):267-72.
22. Goodrich GS, Kabakov AY, Hameed MQ, Dhamne SC, Rosenberg PA, Rotenberg A (2013) Ceftriaxone treatment after traumatic brain injury restores expression of the glutamate transporter, GLT-1, reduces regional gliosis, and reduces post-traumatic seizures in the rat. J Neurotrauma 30(16):1434-41. doi: 10.1089/neu.2012.2712.
23. Götz-Trabert K1, Hauck C, Wagner K, Fauser S, Schulze-Bonhage A. Spread of ictal activity in focal epilepsy. Epilepsia. 2008 Sep;49(9):1594-601. doi: 10.1111/j.1528-1167.2008.01627.x. Epub 2008 Apr 24.
24. Gupta PK, Sayed N, Ding K, Agostini MA, Van Ness PC, Yablon S, Madden C, Mickey B, D'Ambrosio R, Diaz-Arrastia R (2014) Subtypes of post-traumatic epilepsy: clinical, electrophysiological, and imaging features. J Neurotrauma, 31(16):1439-43. doi: 10.1089/neu.2013.3221. Epub 2014 Jul 28.
25. Hameed MQ1, Goodrich GS, Dhamne SC, Amandusson A, Hsieh TH, Mou D, Wang Y, Rotenberg A (2014) A rapid lateral fluid percussion injury rodent model of traumatic brain injury and post-traumatic epilepsy. Neuroreport 25(7):532-6. doi: 10.1097/WNR.0000000000000132.
26. Holmes MD (2008) Dense array EEG: methodology and new hypothesis on epilepsy syndromes. Epilepsia, 49 Suppl 3:3-14. doi: 10.1111/j.1528-1167.2008.01505.x.
27. Hughes JR (1980) Two forms of the 6/sec spike and wave complex. Electroencephalography and Clinical Neurophysiology, 48(5):535-50.
28. Ikeda A, Hirasawa K, Kinoshita M, Hitomi T, Matsumoto R, Mitsueda T, Taki JY, Inouch M, Mikuni N, Hori T, Fukuyama H, Hashimoto N, Shibasaki H, Takahashi R (2009) Negative motor seizure arising from the negative motor area: is it ictal apraxia? Epilepsia 50(9):2072-84.
29. Lüders HO, Engel J, Munari C (1993) General Principles. In: Surgical Treatment of the Epilepsies. Second edition. Editor: J Engel. Raven Press, Ltd, New York. pp. 137-153.
30. Jeha LE, Najm I, Bingaman W, Dinner D, Widdess-Walsh P, Lüders H. Surgical outcome and prognostic factors of frontal lobe epilepsy surgery. Brain 2007;130 (Pt 2):574–84.

31. Kharatishvili I, Nissinen JP, McIntosh TK, Pitkänen A. (2006) A model of posttraumatic epilepsy induced by lateral fluid-percussion brain injury in rats. Neuroscience 140(2):685-97.
32. Marescaux C, Micheletti G, Vergnes M, Depaulis A, Rumbach L, Warter JM (1984) A model of chronic spontaneous petit mal-like seizures in the rat: comparison with pentylenetetrazol-induced seizures. Epilepsia 25(3):326-31.
33. Maroso M, Balosso S, Ravizza T, Iori V, Wright CI, French J, Vezzani A (2011) Interleukin-1β biosynthesis inhibition reduces acute seizures and drug resistant chronic epileptic activity in mice. Neurotherapeutics, 8(2):304-15. doi: 10.1007/s13311-011-0039-z.
34. McIntosh TK, Vink R, Noble L, Yamakami I, Fernyak S, Soares H, Faden AL (1989) Traumatic brain injury in the rat: characterization of a lateral fluid-percussion model. Neuroscience 28(1):233-44.
35. Meeren HK, Pijn JP, Van Luijtelaar EL, Coenen AM, Lopes da Silva FH (2002) Cortical focus drives widespread corticothalamic networks during spontaneous absence seizures in rats. J Neurosci., 22(4):1480-95.
36. Mosewich RK, So EL, O'Brien TJ, et al. Factors predictive of the outcome of frontal lobe epilepsy surgery. Epilepsia 2000;41:843–9.
37. Muro VM and Connolly (2014) Classifying epileptic seizures and the epilepsies. In: Epilepsy. John W. Miller and Howard P. Goodkin, Eds. John Wiley and Sons, Hoboken, NJ. Pp: 10-14.
38. Niedermeyer E, Walker AE, Burton C (1970) The slow spike-wave complex as a correlate of frontal and fronto-temporal post-traumatic epilepsy. Eur Neurol, 3(6):330-46.
39. Norwood BA1, Bumanglag AV, Osculati F, Sbarbati A, Marzola P, Nicolato E, Fabene PF, Sloviter RS (2010) Classic hippocampal sclerosis and hippocampal-onset epilepsy produced by a single "cryptic" episode of focal hippocampal excitation in awake rats. J Comp Neurol. 518(16):3381-407. doi: 10.1002/cne.22406.
40. Paz JT1, Davidson TJ, Frechette ES, Delord B, Parada I, Peng K, Deisseroth K, Huguenard JR (2012). Closed-loop optogenetic control of thalamus as a tool for interrupting seizures after cortical injury. Nat Neurosci. 2013 Jan;16(1):64-70. doi: 10.1038/nn.3269.
41. Pearce PS, Friedman D, Lafrancois JJ, Iyengar SS, Fenton AA, Maclusky NJ, Scharfman HE (2014) Spike-wave discharges in adult Sprague-Dawley rats and their implications for animal models of temporal lobe epilepsy. Epilepsy Behav 32:121-31.
42. Powell KL1, Tang H, Ng C, Guillemain I, Dieuset G, Dezsi G, Çarçak N, Onat F, Martin B, O'Brien TJ, Depaulis A, Jones NC (2014) Seizure expression, behavior, and brain morphology differences in colonies of Genetic Absence Epilepsy Rats from Strasbourg. Epilepsia 55(12):1959-68. doi: 10.1111/epi.12840.
43. Rakhade SN, Klein PM, Huynh T, Hilario-Gomez C, Kosaras B, Rotenberg A, Jensen FE (2011) Development of later life spontaneous seizures in a rodent model of hypoxia-induced neonatal seizures. Epilepsia 52(4):753-65. doi: 10.1111/j.1528-1167.2011.02992.x.
44. Rodgers KM, Bercum FM, McCallum DL, Rudy JW, Frey LC, Johnson KW, Watkins LR, Barth DS (2012) Acute neuroimmune modulation attenuates the development of anxiety-like freezing behavior in an animal model of traumatic brain injury. J Neurotrauma 29(10): 1886-97.
45. Rodgers KM, Dudek FE, Barth DS (2015) Progressive, Seizure-Like, Spike-Wave Discharges Are Common in Both Injured and Uninjured Sprague-Dawley Rats:

Implications for the Fluid Percussion Injury Model of Post-Traumatic Epilepsy. J Neurosci. 2015 Jun 17;35(24):9194-204. doi: 10.1523/JNEUROSCI.0919-15.2015.
46. Roucard C, Pouyatos B, Bouyssieres C, Dumont C, DePaulis A, Duveau V (2014). Effect of the most commonly-used antiepileptic drugs on the MTLE mouse model of human temporal lobe epilepsy: An EEG study. Society for Neurosciences Meeting, November 15-19, Washington, DC, Poster: 521.16/AA1.
47. Semah F, Picot MC, Adam C, Broglin D, Arzimanoglou A, Bazin B, Cavalcanti D, Baulac M (1998) Is the underlying cause of epilepsy a major prognostic factor for recurrence? Neurology, 51(5):1256-62.
48. Shultz SR, Cardamone L, Liu YR, Hogan RE, Maccotta L, Wright DK, Zheng P, Koe A, Gregoire MC, Williams JP, Hicks RJ, Jones NC, Myers DE, O'Brien TJ, Bouilleret V (2013) Can structural or functional changes following traumatic brain injury in the rat predict epileptic outcome? Epilepsia 54(7):1240-50.
49. Stead M, Bower M, Brinkmann BH, Lee K, Marsh WR, Meyer FB, Litt B, Van Gompel J, Worrell GA (2010) Microseizures and the spatiotemporal scales of human partial epilepsy. Brain 133(9):2789-97. doi: 10.1093/brain/awq190.
50. Stefan H and Snead OC (1997) Absence seizures. In: Epilepsy: a comprehensive textbook, (Engel J Jr and Pedley TA eds) pp 579-590. Philadelphia, PA, Lippincott-Raven Publishers.
51. Thompson HJ, Lifshitz J, Marklund N, Grady MS, Graham DI, Hovda DA, McIntosh TK (2005) Lateral fluid percussion brain injury: a 15-year review and evaluation. J Neurotrauma 22(1):42-75
52. Willmore LJ, Sypert GW, Munson JB (1978a) Recurrent seizures induced by cortical iron injection: a model of posttraumatic epilepsy. Ann Neurol. 4(4):329-36.
53. Willmore LJ, Sypert GW, Munson JV, Hurd RW (1978b) Chronic focal epileptiform discharges induced by injection of iron into rat and cat cortex. Science. 200(4349):1501-3.
54. Willoughby JO, Mackenzie L (1992) Nonconvulsive electrocorticographic paroxysms (absence epilepsy) in rat strains. Lab Anim Sci. 42(6):551-4.